# Quantum interference between charge excitation paths in a solid state Mott insulator

S. Wall[1+], D. Brida[2], S. R. Clark[3,1], H.P. Ehrke[1,4], D. Jaksch[1,3], A. Ardavan[1], S. Bonora[2], H. Uemura[5], Y. Takahashi[6], T. Hasegawa[7], H. Okamoto[5,7,8], G. Cerullo[2], A. Cavalleri[1,4]

[1]*Department of Physics, Clarendon Laboratory, Oxford UK*
[2] *IFN-CNR, Dipartimento di Fisica, Politecnico di Milano, Italy*
[3]*Centre for Quantum Technologies, National University of Singapore*
[4]*Max Planck Research Group for Structural Dynamics, University of Hamburg-CFEL*
[5]*Department of Advanced Materials Science, University of Tokyo, Japan*
[6]*Department of Chemistry, Hokkaido University, Sapporo 060-0810, Japan*
[7]*Photonics Research Institute, AIST, Tsukuba 305-8562, Japan*
[8]*CREST-JST, Kawaguchi 332-0012, Japan*

**The competition between electron localization and de-localization in Mott insulators underpins the physics of strongly-correlated electron systems. Photo-excitation, which re-distributes charge between sites, can control this many-body process on the ultrafast timescale[i,ii]. To date, time-resolved studies have been performed in solids in which other degrees of freedom, such as lattice, spin, or orbital excitations[iii,iv,v], come into play. However, the underlying quantum dynamics of 'bare' electronic excitations has remained out of reach. Quantum many-body dynamics have only been detected in the controlled environment of optical lattices[vi,vii] where the dynamics are slower and lattice excitations are absent. By using nearly-single-cycle near-IR pulses, we have measured coherent electronic excitations in the organic salt ET-F$_2$TCNQ, a prototypical one-dimensional Mott Insulator. After photo-excitation, a new resonance appears on the low-energy side of the Mott gap, which oscillates at 25 THz. Time-dependent simulations of the Mott-Hubbard Hamiltonian reproduce the oscillations, showing that electronic delocalization occurs through quantum interference between bound and ionized holon-doublon pairs.**

+ Current address: Dep. of Physical Chemistry, Fritz Haber Institute of the Max Planck Society, Faradayweg 4-6, 14195 Berlin, Germany

In Mott insulators, conductivity at low energies is prevented by repulsion among electrons. This state is fundamentally different from that of conventional band insulators, in which Bragg scattering from the lattice opens gaps in the single particle density of states. The electronic structure of Mott insulators is, therefore, sensitive to doping. Photo-excitation, in analogy to static doping, can trigger large changes in the macroscopic properties[viii]. However, the coherent physics driving these transitions has not been fully observed because the many-body electronic dynamics are determined by hopping and correlation processes that only persist for a few femtoseconds.

We report measurements of coherent many-body dynamics with ultrafast optical spectroscopy in the one-dimensional Mott insulator ET-F$_2$TCNQ. Several factors make this possible: ET-F$_2$TCNQ has a narrow bandwidth (~ 100 meV), which corresponds to hopping times of tens of femtoseconds; the material has a weak electron-lattice interaction; we use a novel optical device producing pulses of 9 fs at the 1.7 μm Mott gap; we study this physics in a one-dimensional system, allowing the evolution of the many-body wavefunction to be calculated and compared with experimental data.

ET-F$_2$TCNQ (bis(ethylendithyo)-tetrathiafulvalene-difluorotetracyanoquinodimethane), is a stacked molecular ionic solid[ix] in which ET molecules are donors and F$_2$TCNQ are acceptors, which form quasi one-dimensional chains of ET molecules. Conducting electrons are localized on ET sites due to the large on-site Coulomb repulsion ($U$ ~ 1 eV) exceeding the single electron hopping amplitude ($t$ ~ 0.1 eV)[x].

Figure 1 displays the optical properties of ET-F$_2$TCNQ. No Drude weight is found at low energies and light polarized parallel to the ET chains is strongly reflected at ~0.7 eV. This feature corresponds to inter-site charge-transfer between neighbouring ET ions, resulting in a hole on one lattice site (a holon) and a neighbouring site with 2 electrons (a doublon). The charge transfer feature in ET-F$_2$TCNQ is sharp, reflecting a bandwidth lower than the gap energy. Figure 1a displays the static reflectivity of ET-F$_2$TCNQ fitted with a multi-Lorentzian dielectric function of the form $\frac{\varepsilon_r(\omega)}{\varepsilon_0} = 1 + \sum_j \frac{A_j}{\omega_{j0}^2 - \omega^2 - i\gamma_j\omega}$ where $A_j$ is the oscillator strength, $\omega_{j0}$ is the resonance frequency and $\gamma_j$ is the damping rate. A single oscillator at $\omega_{Mott}$ = 0.675 eV is sufficient to describe the Mott gap. Figure 1b shows the optical conductivity obtained from the imaginary part of the fitted dielectric function.

Photo-excitation across the Mott gap of ET-F$_2$TCNQ transiently generates a metallic state[xi], evidenced by the transfer of spectral weight from the Mott gap to a Drude response at low frequencies resulting from charge delocalization. To observe the initial delocalization dynamics, short light pulses are needed. Even for the low hopping amplitudes of this compound ($t$ ~ 100 meV) we expect electron delocalization to require a time of order $h/t$ ~ 40 fs, where $h$ is Planck's constant. Similarly, we expect correlated electrons to be dressed on timescales of the order of $h/U$. Furthermore, the pulses need to be resonant with the 0.7 eV charge transfer resonance, corresponding to a wavelength of 1.7 μm, where sub-10-fs pulses have not been previously achieved. Thus, we developed a source at 1.7 μm with 9 fs duration[xii].

Figure 2 reports time-resolved reflectivity measurements of the photo-induced response of ET-F$_2$TCNQ in the paramagnetic, Mott insulating phase at room temperature. The time-dependent reflectivity was probed with a replica of the pump pulse and spectrally resolved. Both pump and probe were polarized along the crystal's *a*-axis, to excite charges across the Mott gap. Figure 2a shows the spectrally-integrated reflectivity change. After a prompt decrease, relaxation back to the ground state occurs with a bi-exponential decay with 130 fs and 840 fs time constants. Excitation with light perpendicular to the chains produced no observable dynamics, confirming the experiments address the one-dimensional Mott physics of ET-F$_2$TCNQ.

Figure 2b shows the spectrally resolved reflectivity. A prompt shift of spectral weight towards lower energies is followed by a drop in the reflectivity at the Mott gap. The dashed lines in Fig. 2 show the fitted reflectivity at each delay. Whilst the unperturbed reflectivity was accurately described with a single oscillator at the Mott gap, the transient reflectivity of the photo-excited system requires an additional, lower energy, oscillator. The appearance of a new resonance, which is clearly visible in the reflectivity lineouts of Fig. 2, suggests that below-gap bound states are formed, reminiscent of bound holon-doublon pairs observed in other Mott insulators[xiii].

The characteristics of this new peak are time dependent, as visualized in Fig. 2c, where we have normalized the reflectivity at each time step. Two contours are shown in Fig. 2c. On the blue side, a prompt red-shift and recovery of the resonance is observed, whereas the red side shows a longer-lived component, containing a damped oscillatory response at 25 THz. Static

Raman data on ET-F$_2$TCNQ does not show any equivalent features, strongly suggesting that the oscillation is not due to coherent phonons, but of an electronic origin[xiv].

To investigate such dynamics, we used a one-dimensional Mott-Hubbard Hamiltonian for a half-filled chain, with $N$ = 10 sites, with electron hopping, $t$, and onsite and nearest neighbour Coulomb repulsion $U$ and $V$,

$$H = -t \sum_{l,\sigma}^{L} \left( c_{l,\sigma}^{\dagger} c_{l+1,\sigma} + c_{l+1,\sigma}^{\dagger} c_{l,\sigma} \right) + U \sum_{l} n_{l,\uparrow} n_{l,\downarrow} + V \sum_{l} n_l n_{l+1}$$

where, $c^{\dagger}_{l,\sigma}$ and $c_{l,\sigma}$ are the creation and annihilation operators for an electron at site $l$ with spin $\sigma$, $n_{l,\sigma}$ is the number operator, and $n_l = n_{l,\uparrow} + n_{l,\downarrow}$. We described the initial state as $\rho_g = \frac{1}{2^N} \sum_{\underline{\sigma}} |\psi_{\underline{\sigma}}\rangle\langle\psi_{\underline{\sigma}}|$, where $\psi_{\underline{\sigma}}$ represents a many-body wavefunction with one electron per site and total spin-vector $\underline{\sigma}$. This reflects the fact that, at room temperature, charges are localized, but posses no magnetic ordering.

We calculate the static optical conductivity (see methods section) to find values of $U$, $V$ and $t$ that provide the best fit to the experimental results. The best fit, shown in Fig. 3c ($t$ = -200 fs), gave $U$ = 820 meV, $V$ = 100 meV and $t$ = 50 meV. It was not possible to fit the optical conductivity using $U$ and $t$ alone and inter-site correlation energy, $V$, was needed[xv].

These static parameters were used to fit to the time-dependent optical properties. We considered states created by the laser, $\rho_{e,L}$, which consist of a neighbouring holon-doublon pair, delocalized over $L$ lattice sites, with an optical conductivity $\sigma_{e,L}(\omega)$. To model the time-dependence, the system was described by an incoherent mixture of two states: $\rho(T)$ = $p_g(T)\rho_g + p_e(T)\rho_{e,1}$, where $T$ is pump-probe time delay and $p_s(T)$ is the fraction of the sample in state $s$. The time-dependent optical conductivity was described as $\sigma(\omega, T) = p_g(T) \sigma_g(\omega) + p_e(T) \sigma_{e,1}(\omega)$, and $p_g(T)$ and $p_e(T)$ were varied, under the constraint $p_g(T) + p_e(T) = 1$, to obtain the best fit.

The result is displayed in Fig. 3b, where the calculated time-dependent optical conductivity is compared to that obtained from the transient reflectivity[xvi]. The model provides a good fit at long time delays (>200 fs), indicating that the long-term dynamics are dictated by incoherent holon-doublon decay. However, the model fails at early times when coherent processes are primarily responsible for the dynamics.

To simulate the coherent dynamics, we calculate the time evolution of the state $\rho_{e,10}$ under the Hamiltonian. The temporal evolution has no free parameters, and the optical conductivity was calculated after the state evolved for a time $T$. The normalized result of this simulation is compared to the normalized experimental optical conductivity in Fig. 4. The simulation reproduces the 25 THz oscillations observed experimentally, as shown by the Fourier transforms in Fig. 4c. The numerical simulation contains no damping or de-phasing, resulting in a narrower Fourier transform, yet, it is remarkable that this model reproduces the frequency with this accuracy.

The origin of these oscillations can be understood by considering interference between different photo-excitation paths of the Mott insulator. Optical excitation acts on the ground state |…111111…> to create bound states of the form |…11**0**2**11…>. In this case, the holon-doublon pair is bound by an energy $V$. This state can evolve into a superposition of bound states and ionized states of the form |…11**0**12**1…> via electron hopping. These excitations interfere in the time domain, giving rise to the observed oscillations. This conclusion is validated by simulations with $V = 0$, in which no oscillations are observed (see supporting online material).

This physics is reminiscent of single particle excitations in multiple quantum wells. An electron is transferred from a well of depth $U$ into a neighbouring well with depth $V$, representing the bound holon-doublon pair. This 'exciton' can tunnel, with an energy $t$, into a third well, representing the ionized pair (see Fig. 4d). Such a system will oscillate at $\Omega = \sqrt{(V^2 + t^2)}/h$ = 27 THz, in close agreement with the experimental and numerical observations. Our experiments highlight, in a room temperature solid, coherent many-body physics that has, so far, only been accessible in ultracold gases[xvii]. Our experiments captured dynamics on the timescale associated with hopping and inter-site correlations. However, higher temporal resolution could observe dressing due to onsite Coulomb correlations, and the coherent formation of the holon-doublon pairs. These measurements would require temporal resolutions approaching the attosecond regime[xviii], therefore, we anticipate that time-resolved photo-electron spectroscopy[xix] and ultrafast soft x-ray techniques[xx] will play major roles in these studies.

**Acknowledgements**
H.O. is grateful for support by a Grant-in-Aid for Scientific Research (No. 20110005) by the Ministry of Education, Culture, Sports, Science and Technology of Japan.

**Correspondence**

Correspondence and requests for material should be sent to Simon Wall, wall@fhi-berlin.mpg.de or Andrea Cavalleri, andrea.cavalleri@mpsd.cfel.de

**Author Contributions**

S.W, D.B, H.P.E and G.C performed pump-probe experiments. D.B, S.B and G.C designed and built the experimental apparatus. H.U, Y.T, T.H and H.O provided samples. S.W analyzed the experimental data. S.R.C and D.J performed the numerical simulations. S.W, A.C, A.A, S.R.C and D.J interpreted the data and the simulations. The manuscript was written by A.C and S.W. A.C conceived and coordinated the project.

**Methods**

**Sample preparation**

Single crystals were grown by first purifying commercially available ET molecules and synthesizing $F_2TCNQ$[xxi] by repeated cycles of re-crystallization and sublimation. Single crystals of approximately 3mm×10mm×0.5 mm were achieved by slowly cooling the hot chlorobenzene solution of purified ET and $F_2TCNQ$.

**Experimental setup**

A near-IR Optical Parametric Amplifier (OPA) is driven by an amplified Ti:Sapphire laser (100 μJ, 150 fs, 1 kHz pulses at 800 nm). A white-light seed, generated in a sapphire plate, is amplified in a 3 mm Type-I β-barium-borate crystal to ≈2 μJ energy. The ultrabroadband OPA pulses, with spectrum covering the 1200-2200 nm range, are compressed using a deformable mirror to nearly transform-limited 9 fs duration. The pump fluence at the sample surface was 3.5 mJ/cm$^2$. Transient reflectivity is measured in a degenerate pump-probe configuration at a near-normal angle of incidence and the probe spectrum was detected by an InGaAs optical multichannel analyzer.

**Numerical simulation**

The initial state of the system is assumed to be described by the density matrix, $\rho_g = \frac{1}{2^N} \sum_\sigma |\psi_\sigma\rangle\langle\psi_\sigma|$. The laser creates excited states of the form $\rho_{e,L} \propto X_L \rho_g X^\dagger_L$, where the excitation operator $X_L = \sum_{\sigma,l=1}^{L} \left( c^\dagger_{l,\sigma} c_{l+1,\sigma} - c^\dagger_{l+1,\sigma} c_{l,\sigma} \right)$ corresponds to the creation of a single holon-doublon pair at neighbouring sites, delocalized over $L$ lattice sites.

The optical conductivity of a given state, $\rho$, is calculated via the unequal time current-current correlation function $c_{jj}(\tau,\tau') = \text{tr}[\rho j(\tau) j(\tau')]\theta(\tau-\tau')$, where $\tau > \tau'$, $j = it \sum_{l,\sigma} \left( c_{l,\sigma}^\dagger c_{l+1,\sigma} - c_{l+1,\sigma}^\dagger c_{l,\sigma} \right)$ is the current density operator, $\tau$ is time, $\theta(\tau)$ is the Heaviside function and $j(\tau) = \exp(iH\tau) j \exp(-iH\tau)$ is the current operator in the Heisenberg picture. The definition of *H* is given in the main text.

For the incoherent model we take $c_{jj}(\tau,\tau') = c_{jj}(\tau,0) = c_{jj}(\tau)$ and the regular finite-frequency optical conductivity then follows from $c_{jj}(\omega)$, the Fourier transform taken with respect to $\tau$, as $\sigma(\omega > 0) \propto \frac{\text{Re}\{c_{jj}(\omega)\}}{\omega}$. The total evolution time over which $c_j(\tau)$ is computed was limited to $\tau_{max} = 5\hbar/t$. The Fourier transform was performed with a Gaussian windowing function $exp[-4(\tau/\tau_{max})^2]$, leading to a broadening and smoothing of $c_{jj}(\omega)$ compared to its exact limit, which is instead composed of numerous $\delta$-functions. This is justified because we focus on features in the high angular frequency range $8t < \omega < 32t$ and because the probe pulse used in the experiment has similar spectral limitations.

The time-dependent optical conductivity for the state $\rho_{e10}$ was calculated using the full two-time current-current correlation function $c_{jj}(\tau,T)$, where *T* is the pump-probe time delay and the Fourier transform is taken with respect to $\tau-T$.

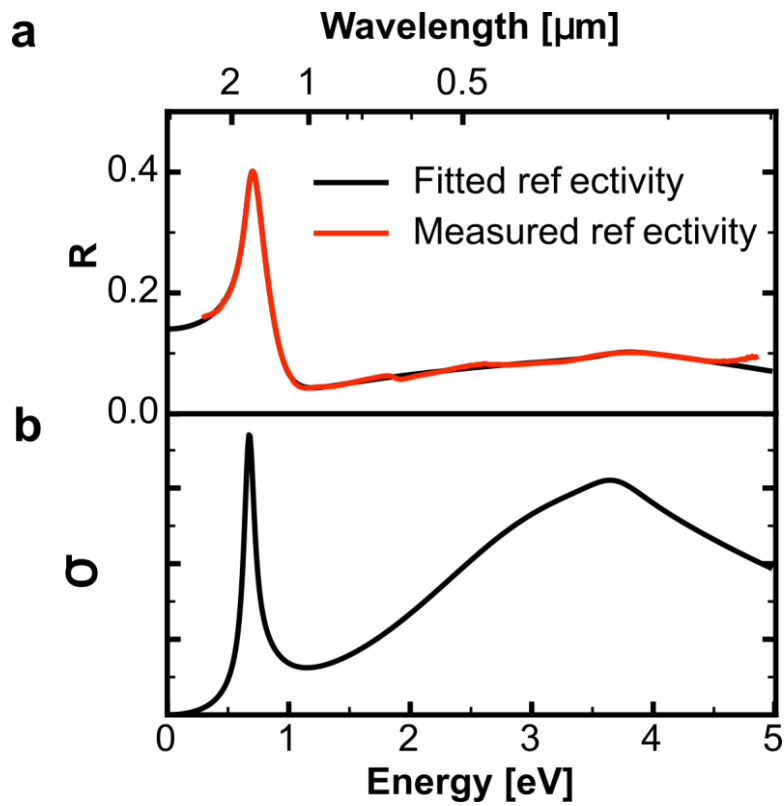

**Figure 1 Steady state reflectivity and conductivity of ET-F$_2$TCNQ. (a)** Measured reflectivity (red line) and modelled reflectivity (black line). The reflectivity was obtained from the dielectric function described in the text with $R = \left|\frac{1-\sqrt{\varepsilon_r}}{1+\sqrt{\varepsilon_r}}\right|$ **(b)** optical conductivity extracted from the imaginary part of the dielectric function.

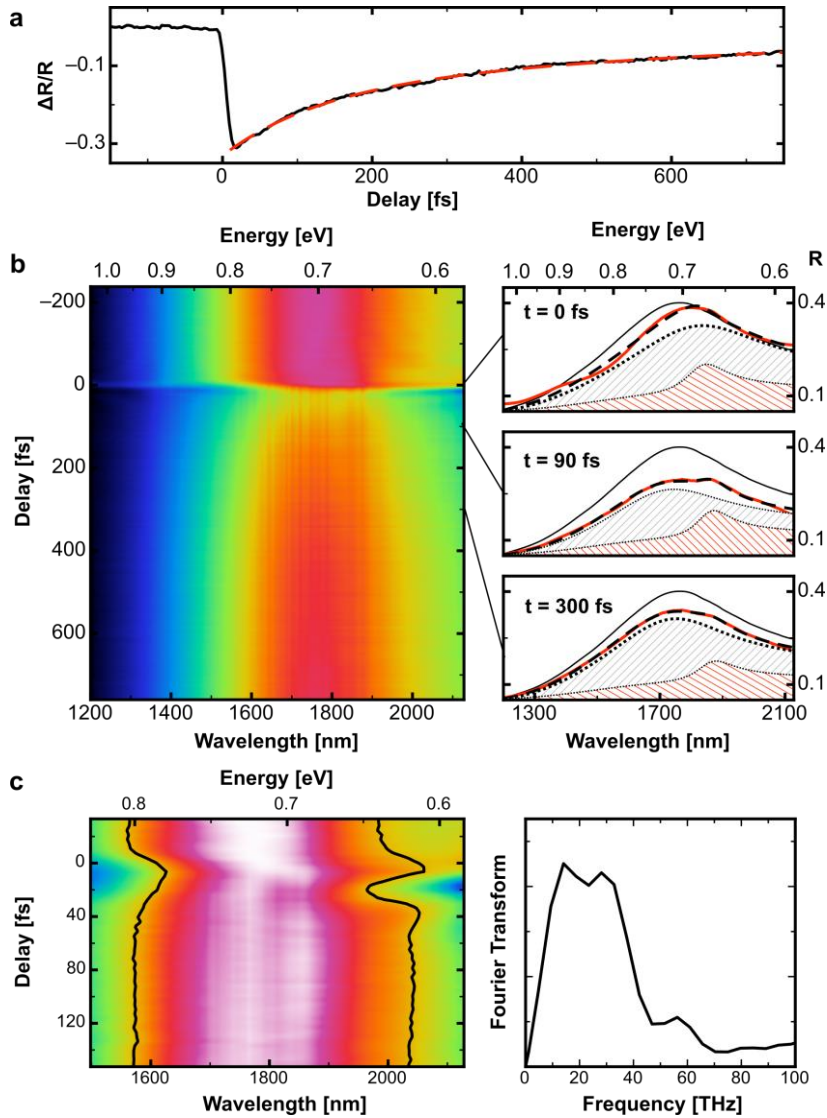

**Figure 2 Transient reflectivity of ET-F$_2$TCNQ. (a)** Spectrally integrated, time-dependent optical reflectivity of ET-F$_2$TCNQ with a bi-exponential fit (dashed red curve). The resonance at the Mott gap collapses and then recovers with two time constants of 130 fs and 840 fs. **(b)** Two-dimensional colour plot of the time-dependent reflectivity for the spectral region between 1.2 μm and 2.1 μm. The lineouts show the evolution of the reflectivity (red line) together with the fitted reflectivity (dashed black line) compared to the static reflectivity (thin black line). The shaded lines represent the contribution to the reflectivity from the Lorentzian functions used to fit the peaks. **(c)** Normalized reflectivity at early times, showing oscillations on the red edge of the spectrum. The right hand side panel shows a Fourier transform of the oscillations in the reflectivity at 2 μm.

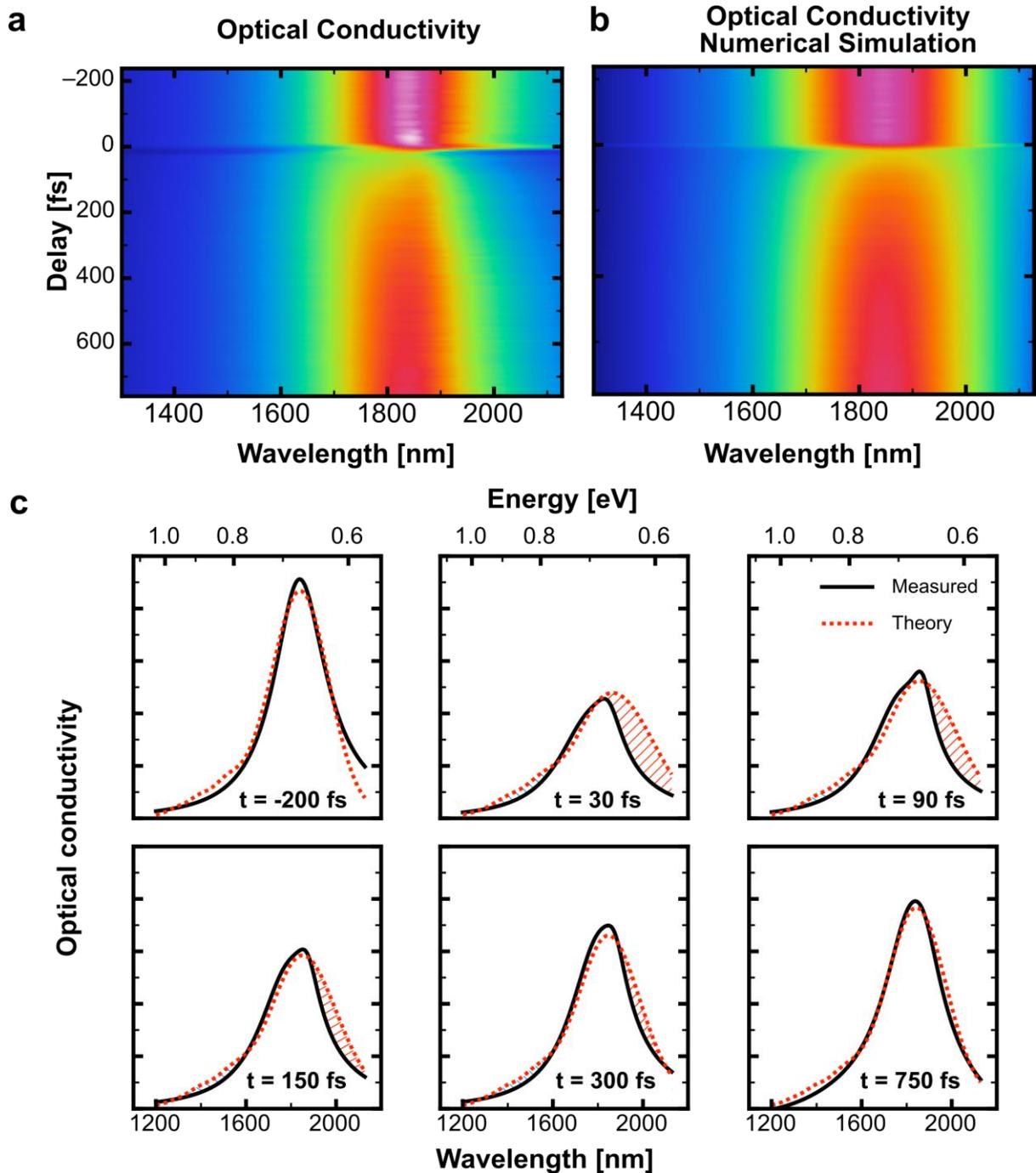

**Figure 3 Retrieved and simulated optical conductivity.** (a) Two-dimensional colour plot of the time and wavelength-dependent optical conductivity extracted from the reflectivity data. (b) Numerical simulation of the optical conductivity with $\sigma(\omega, T) = p_g(T)\,\sigma_g(\omega) + p_e(T)\,\sigma_{e,1}(\omega)$ (c) Comparison of the numerical model to the extracted optical conductivity. The model is in increasing agreement with measurements for time delays greater than 200 fs, indicating that the recovery of the system is due to the decay of holon-doublon pairs.

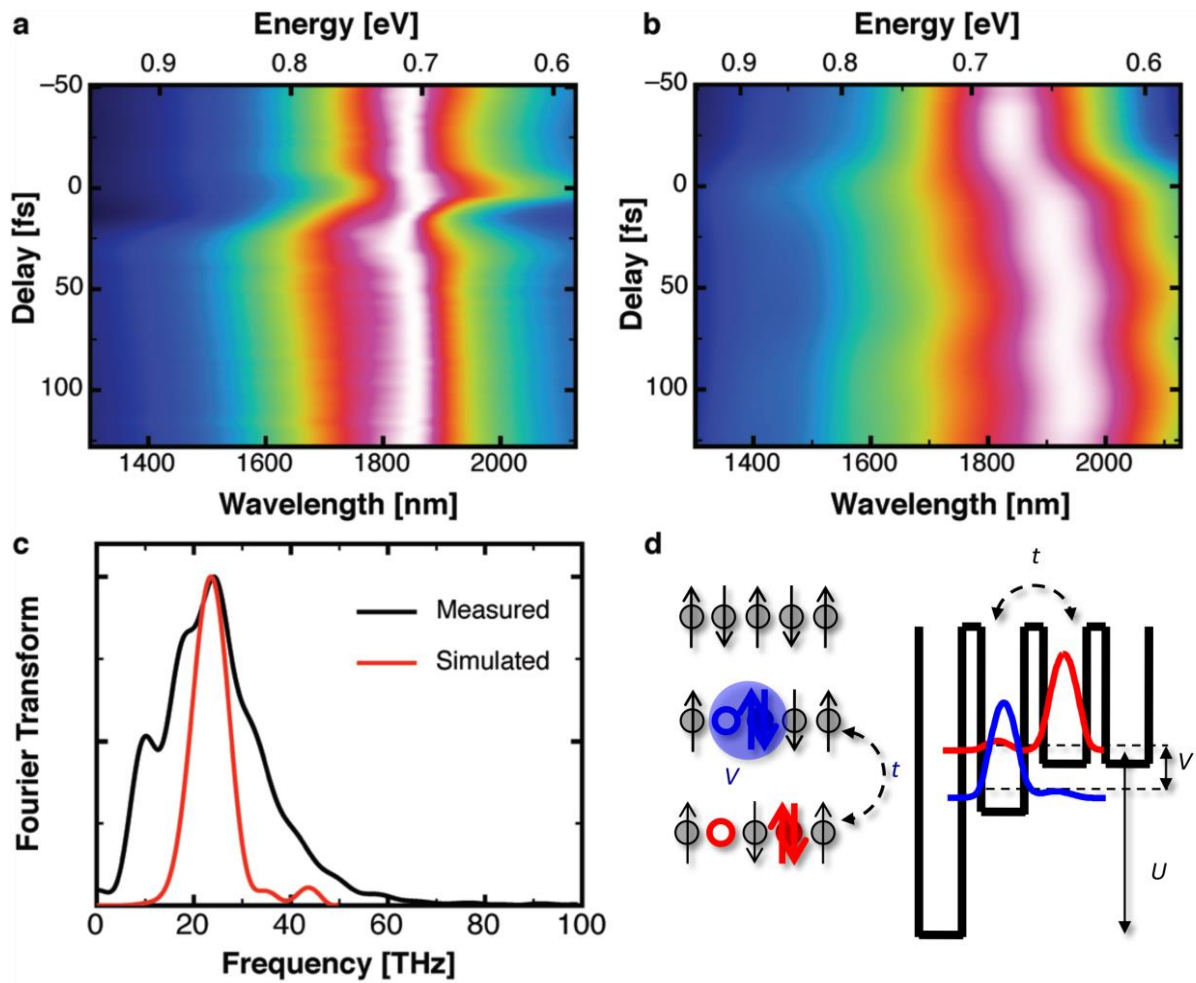

**Figure 4 Coherent oscillations of optical conductivity. (a)** The retrieved normalized optical conductivity during the first 150 fs, with oscillations in spectral weight with a period of approximately 40 fs. **(b)** Time dependent simulation of the quantum evolution of the optical conductivity under the Mott-Hubbard Hamiltonian showing similar oscillations. **(c)** Fourier transforms of the oscillations in (a) and (b), showing an excellent agreement at ~25 THz. **(d)** Representations of the initial state, excited holon-doublon bound state, and ionized holon-doublon state. (Right) Mapping the bound and ionized holon-doublon states onto two potential energy wells offset by $V$ and coupled by tunnelling amplitude $t$.

# Supporting Online Information

Here we expand the discussion of the interpretation of the observed oscillations from both experimental and theoretical perspectives.

## Raman measurements

As discussed in the main body of the letter, oscillations in time resolved reflectivity measurements can often result from the impulsive generation of coherent phonons by the pump laser pulse. Here, we discuss how this can be ruled out by comparing the observed oscillations to the Raman scattering signal of ET-F$_2$TCNQ.

Raman scattering measurements were performed using a 633-nm-wavelength laser. Experiments were performed in a backscattering geometry at room temperature. Figure S1 shows the Raman signal of ET-F$_2$TCNQ for different polarizations within the *ab* plane. For all polarizations, the Raman spectra consist of a series of narrow lines, corresponding to long-lived vibrations in the time domain. No polarization combination gives rise to significant peaks in the region from ~700-1100 cm$^{-1}$, which could correspond to the observed ~25 THz (830 cm$^{-1}$) oscillation in the optical conductivity.

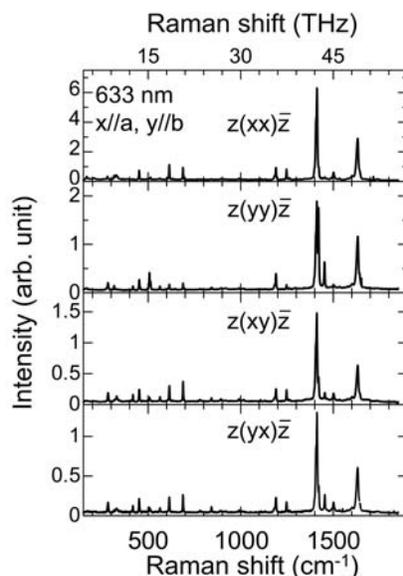

**Figure S1** Raman back-scattering measurements on ET-F$_2$TCNQ within the *ab*-plane at room temperature. Light is incident along the *z* axis (at a small angle to the crystal's *c* axis). The incident and detection polarization directions are indicated in brackets. The (xx) configuration has the closest correspondence to the time resolved experimental setup.

In addition to this, the majority of these modes can be excited equally with light polarized along either the *a* or the *b* crystal axis, showing that these modes do not follow the one dimensional nature of the electronic structure. If the observed oscillation in the optical conductivity was a result of such a phonon, it should be excitable when the sample is pumped along the *b*-axis, however, no transient signal was observed above the noise in this configuration. Therefore, we conclude that time-domain oscillations cannot be assigned to coherent phonons.

### *Theoretical simulations*

In the manuscript we report that oscillations found in the numerical simulations of the time dependent optical conductivity of ET-F$_2$TCNQ, and use a simple model to suggest that this dynamic is the result of coherent oscillations between bound and ionised holon-doublon pairs. In this section, we present further numerical calculations to justify this claim.

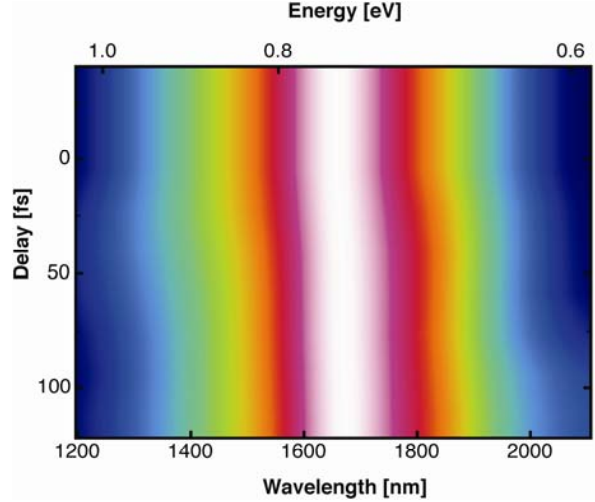

**Figure S2** Normalized time-dependent optical conductivity calculated with *V = 0*.

Figure S2 shows the normalized, time-dependent optical conductivity when *V = 0*, i.e. when no inter-site interaction is present and thus when neighbouring and separated pairs have degenerate energies. This shifts the initial state peak position to higher energies. However, most importantly, unlike figure 4b of the main text, which has a finite *V*, high frequency oscillations are not observed, in agreement with the model presented in the text[1].

The results of the simulation with *V = 0* demonstrate that the holon-doublon interaction is *necessary* to produce the observed high frequency oscillations in the numerical data. However, in order to show that the oscillation is due to interference between bound and ionized holon-doublon states we also calculate the time-dependent expectation value for the number of bound-holon pairs for the Hamiltonian given in the text, as

$$\langle hd \rangle = \sum_{i=1}^{N}(h_i d_{i+1} + h_i d_{i-1}),$$

where $d_i = n_{i\uparrow}n_{i\downarrow}$ and $h_i = (1 - n_{i\uparrow})(1 - n_{i\downarrow})$ are the doublon and holon number operators respectively at site *i*.

---

[1] At longer times, for both *V = 0* and for finite *V*, lower frequency oscillations in the optical conductivity at ~7 THz, of order h/t, are observed. Rapid de-phasing of the experimental data at room temperature prevents the observation of such oscillations.

Figure S3a shows the temporal evolution of the number of bound pairs in the system. Clear oscillations are observed on top of a decaying background. The Fourier transform of these oscillations, shown in figure S3b, peaks at the same value observed in the numerical simulations of the optical conductivity. Examining the behaviour of correlations between holon-doublon pairs beyond nearest neighbours indicates that the decay of the correlation is likely to be due to diffusion of the ionized pairs and additional effects which will be the subject of further investigations. Nonetheless, the results presented clearly prove that observed oscillations in the optical conductivity are predominantly a consequence of interference between bound-ionized holon-doublon states.

a)                                                         b)

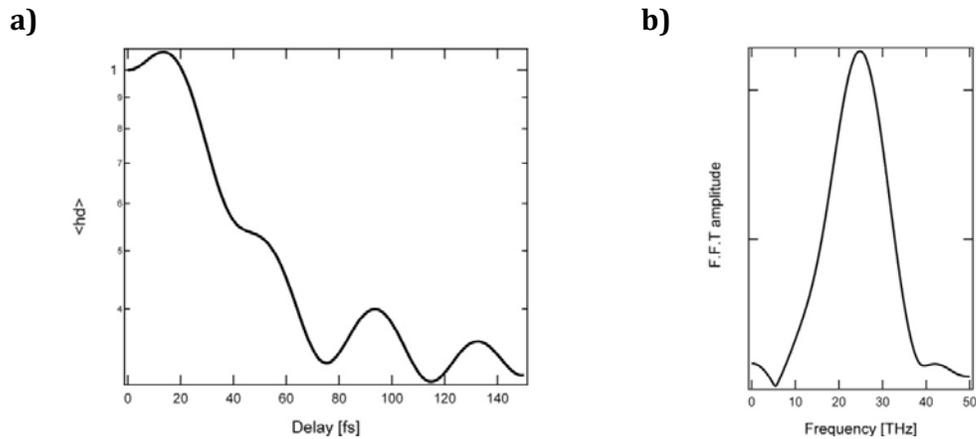

**Figure S3 a)** Time dependence of the number of bound holon doublon pairs. **b)** Fourier transform of the observed oscillations in a) after a background subtraction.